\documentclass[ 
 superscriptaddress,
 amsmath,amssymb,
 aps,
 twocolumn,
 pra,
]{revtex4-2}
\usepackage{color}
\usepackage{comment}
\usepackage{graphicx}
\usepackage{dcolumn}
\usepackage{bm}
\usepackage{hyperref}

\begin{document}

\preprint{APS/123-QED}

\title{Observation of 2D dam break flow and a gaseous phase of solitons in a photon fluid}

\author{Ludovica Dieli}
\email{ludovica.dieli@uniroma1.it}
\affiliation{Department of Physics, Sapienza University, 00185 Rome, Italy}
\affiliation{Enrico Fermi Research Center (CREF), 00184 Rome, Italy}
\author{Davide Pierangeli}
\affiliation{Institute for Complex Systems, National Research Council, 00185 Rome, Italy}
\author{Eugenio DelRe}
\affiliation{Department of Physics, Sapienza University, 00185 Rome, Italy}
\author{Claudio Conti}
\affiliation{Department of Physics, Sapienza University, 00185 Rome, Italy}
\date{\today}

\begin{abstract}
We report the observation of a two-dimensional dam break flow of a photon fluid in a nonlinear optical crystal. By precisely shaping the amplitude and phase of the input wave, we investigate the transition from one-dimensional (1D) to two-dimensional (2D) nonlinear dynamics. 
We observe wave breaking in both transverse spatial dimensions with characteristic timescales determined by the aspect ratio of the input box-shaped field. The interaction of dispersive shock waves propagating in orthogonal directions gives rise to a 2D ensemble of solitons.
Depending on the box size, we report the evidence of a dynamic phase characterized by a constant number of solitons, resembling a 1D solitons gas in integrable systems.
We measure the statistical features of this gaseous-like phase.
Our findings pave the way to the investigation of collective solitonic phenomena in two dimensions, 
demonstrating that the loss of integrability does not disrupt the dominant phenomenology.
\end{abstract}

\maketitle

Optics in nonlinear media provides a unique field
to investigate nonlinear statistical phenomena such as dispersive shock waves (DSWs)~\cite{el2016dispersive},
Fermi-Pasta-Ulam recurrences~\cite{pierangeli2018observation, kimmoun2016modulation, mussot2018fibre}, chaos~\cite{xin2021evidence}, and replica symmetry breaking~\cite{pierangeli2017observation}.
The study of the collective dynamics of nonlinear waves, such as the formation of a soliton gas (SG)
~\cite{redor2019experimental, marcucci2019topological, suret2020nonlinear, fache2023interaction, suret2023gas}, is a frontier with significant applications from supercontinuum generation to neuromorphic computing~\cite{marcucci2020theory, silva2021reservoir, lopez2023self}. However, experimental investigations have been limited to one-dimensional (1D) waves.
The difficulties in high-dimensional experiments are due to the collapse and the modulation instability that dominates the generation of soliton ensembles. 
Theoretical results are also limited by the lack of integrability.
Nevertheless, the study of hydrodynamic regimes in quantum photon fluids~\cite{carusotto2013quantum, wan2007dispersive, carusotto2014superfluid, situ2020dynamics, xu2015coherent, xu2017dispersive, conti2009observation, bienaime2021quantitative, bonnefoy2020modulational, audo2018experimental, bendahmane2022piston} 
triggers intense research on the complex dynamics of solitons in the framework of the nonlinear Schroedinger equation (NLS).  The NLS is well approximated by hydrodynamic models with a quantum pressure, which distinguishes classical from quantum fluids~\cite{pitaevskii2016bose}.

In this context, the dam break flow is a paradigm for the emergence of highly nonlinear phenomena.
Theoretical works~\cite{el2016dam, biondini2018riemann} have shown that the asymptotic state
of a 1D dam break flow is a SG. The SG, introduced by Zakharov~\cite{zakharov1971kinetic}, 
is an ensemble of interacting solitons with a random distribution of phase and amplitude~\cite{el2021soliton, suret2023soliton}. Integrability is fundamental for the definition of SG concept. 
For this reason, to date, experimental evidence of a SG has been reported only in hydrodynamic or nonlinear optical platforms that approach a 1D system~\cite{redor2019experimental, marcucci2019topological, suret2020nonlinear, fache2023interaction, suret2023gas}.

In this Letter, we report the first experimental observation of a two-dimensional (2D) dam break flow.
An engineered 2D coherent optical wave undergoes the dam break by propagating in a photorefractive crystal.
The complete control of the input waveform gives access to the transition from the 1D to the 2D regime, unveiling the crucial role of the second spatial dimension in the wave-breaking. We exploit the multi-dimensional interaction of 1D shocks colliding in orthogonal directions to trigger a soliton-generating transition. 
We reveal two nonlinear scales: the first regards the 1D wave breaking, the second the collision of DSWs and the evolution into a 2D ensemble of solitons. The agreement with the soliton existence curve proves the solitonic nature of the observed localized modes. 
We observe a steady state with a well-defined number of solitons, dependent on the box size, 
suggesting the presence of a gaseous phase of solitons~\cite{schwache1997properties}. 
This is an unexpected result because the SG concept is strictly related to an integrable nonlinear model, while in non-integrable dynamics the expected asymptotic state is the coalescence of the soliton ensemble. 
Hence, our observations raise the question if a definition of a 2D SG is possible.
Despite our experiments being far from integrable regimes, we report a nearly thermalized state in which the soliton number is conserved and the intensity probability distribution function (PDF) 
follows an exponential law compatible with the statistics of a 1D SG.
Our evidence of a dynamical phase of solitons with gaseous features can stimulate further experimental and theoretical research.

\begin{figure}
    \centering
    \includegraphics[width=\linewidth]{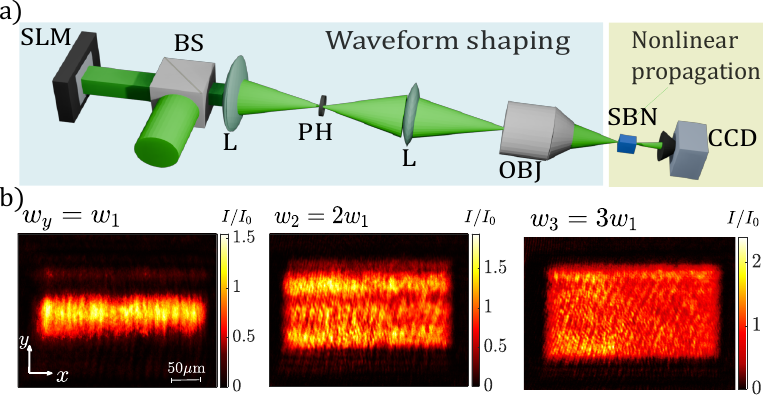}
    \caption{Coherent shaping of the 2D input wave. 
    (a) Experimental setup. The laser beam is shaped in amplitude and phase by a system made by a SLM, two lenses (L), a pinhole (PH), and a de-magnification objective (OBJ). The beam propagates in a biased photorefractive crystal (SBN) and the output intensity is detected by a camera (CCD).
        (b) Intensity distributions for three box-shaped initial conditions of different heights $w_y$.
     $I_0$ is the mean intensity over the box.}
    \label{fig:setup}
\end{figure}

A sketch of our setup is reported in Fig.\ref{fig:setup}(a). 
The  wavelength of the continuous-wave laser is $\lambda=532\,\text{nm}$. 
We implement a novel waveform shaping system for the preparation of the initial condition, 
composed of a phase-only spatial light modulator (SLM),
a filter in Fourier space made by two bi-convex lenses in 4-$f$ configuration ($100\,\text{mm}$ focal length) with a precision pinhole ($500\,\mu \text{m}$ diameter), 
and a large-area objective ($0.1$ magnification ratio) to demagnify the beam. 
Beam propagation ($10\, \text{mW}$ power) occurs in a photorefractive crystal 
of SBN ($\text{Sr}_\text{x} \, \text{Ba}_{(1-\text{x})}\,\text{Nb}_2\text{O}_6$, with $\text{x} = 0.61)$ of size $5\times 5\times 5\,\text{mm}^3$, 
where the nonlinearity is induced by an external electrostatic field via a high-voltage power supply.
The optical field at the crystal output facet is imaged onto an intensity camera.
Waveform shaping is based on the interferometric method reported in Refs.~\cite{mendoza2014encoding,carbonell2019encoding},
 enabling the encoding of a complex field by a single phase-only SLM. 
The technique allows us to generate a 2D coherent beam of arbitrary shape.
We obtain a fully programmable initial wave with micrometric spatial resolution.
This solves a challenging problem in wave optics, enabling complete control of nonlinear optical propagation in crystals.

Exploiting the programmability of the input waveform, we implement a box-shaped initial condition whose side lengths can be independently varied. This allows us to explore the role of dimensionality in the dam break flow and the transition from a 1D to a 2D regime.
We consider as input waves three boxes of width $w_x=300\,\mu\text{m}$ and different heights $w_{y}$:
 $w_1=50\,\mu \text{m}$, $w_2=100\,\mu \text{m}$ and $w_3=150\,\mu \text{m}$. The box shape is generated as a supergaussian beam with a flat phase (see Supplementary Material \cite{suppli}). The experimental realization of this initial condition [Fig.\ref{fig:setup}(b)] presents intensity fluctuations due to inhomogeneities in beam shaping. Intensity fluctuations have been included in our numerical simulations as an additive white noise $\eta(x,y)$. \\
As detailed in the Supplementary Material~\cite{suppli}, the nonlinear response of the photorefractive crystal is due to a space charge field generated by the combined effect of the external field and the impinging beam intensity. This process produces a time-dependent variation of the refractive index $n(I,t)=n_0+\delta n(I,t)$. 
For tetragonal SBN, the nonlinear index variation is $\delta n(I,t)=\delta n_0\frac{f(t)}{\left(1+\frac{I}{I_s}\right)} $ where $\delta n_0$ contains the external field and the linear electro-optic coefficients. $I_s$ is the saturation intensity related to the background illumination, and $f(t)$ is the charge accumulation function \cite{delre2009photorefractive}.
The nonlinear propagation of the optical field is described by a generalized 2D NLS equation
 with a focusing Kerr-saturable nonlinearity 
\begin{equation}
    2 {\textrm i} k\, \partial_{z}A+\partial_{x}^{2}A+\partial_{y}^{2}A+2 k^{2}\frac{\delta n(I,t)}{n_0} A=0
    \label{eq:NLS_kerrsat}
\end{equation}
where $k$ is the wavenumber.
In the experiments, the beam is observed at the crystal output ($z=5\,\text{mm}$, applied voltage $800$ V)
 as a function of time $t$.
According to Eq.(\ref{eq:NLS_kerrsat}), the spatial propagation in the $z$-direction
can be mapped into a time evolution at a fixed $z$-distance by a dimensionless 
time-dependent propagation variable \cite{marcucci2019topological} (see Supplementary Material~\cite{suppli}).
We support experimental observations with numerical simulations. Numerical results are obtained from
Eq.(\ref{eq:NLS_kerrsat}) in dimensionless form for a box-shaped initial condition with noise $|\eta| \ll A_{0}$. Simulations provide the beam propagation in space at a fixed evolution time (nonlinear coefficient).

\begin{figure*}[t]
    \centering
    \includegraphics[width=\linewidth]{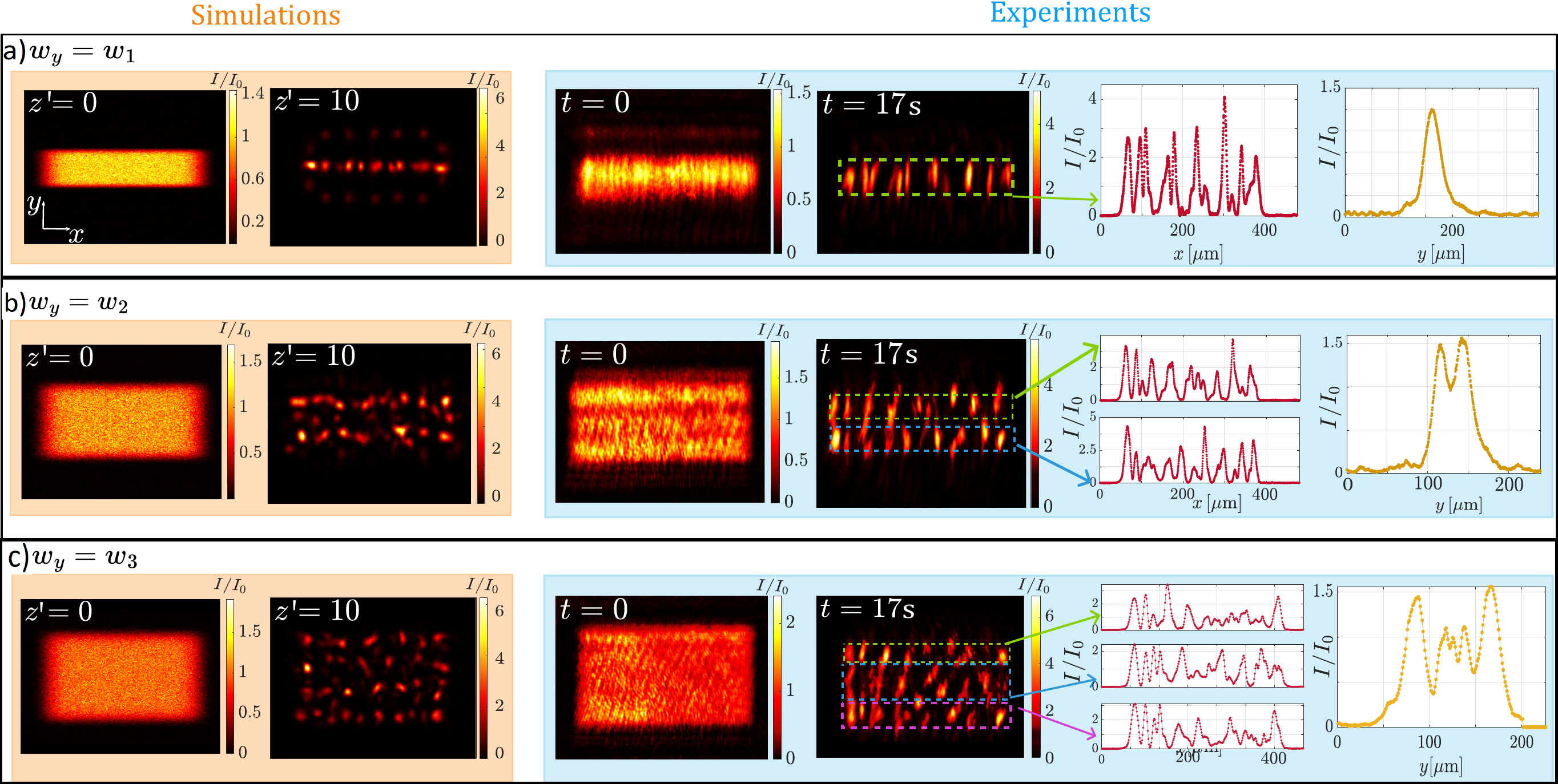}
    \caption{Two-dimensional dam break flow of a photon fluid. Input and output intensity distributions
 for boxes with heights: a) $w_1=50\,\mu\text{m}$, b) $w_2=100\,\mu \text{m}$, c) $w_3=150\,\mu\text{m}$. Intensity is normalized to the mean input intensity $I_0$. 
     Numerical results at $z'=10$ closely match observations at the crystal output ($z=5$ mm) for $t=17\,\text{s}$. Experimental intensity profiles along $x$ and $y$ are averaged over the dotted boxes.} 
    \label{fig:evolution}
\end{figure*}

Figure \ref{fig:evolution} shows the observed and simulated evolution of the three input boxes in Fig.\ref{fig:setup}(b). The boxes break up in a multiple-soliton structure. The agreement between the measured and simulated intensity is found at the dimensionless distance $z^\prime=10$, validating the matching between the model and the experiment. 
Fig.\ref{fig:evolution}(a) reports the results for the thin stripe ($w_y=w_1$) mapping the 1D system. 
As expected for the dam break problem~\cite{el2016dam}, the breaking occurs along the $x$-axis.
The two input discontinuities form a pair of DSWs that collide generating a 1D SG after $t=17$ s.
Increasing $w_y$ causes an additional breaking along the $y$-axis.
We observe the formation of an ensemble of 2D solitons with random intensities that are distributed in space on multiple rows [Fig.\ref{fig:evolution}(b)-(c)].
The height of the box is crucial in governing the 2D breaking process.
When $w_y$ is small ($w_y\simeq 50\,\mu \text{m}$) the dynamics is 1D, while, above a threshold height, the long-time evolution shows the formation of soliton ensemble occupying the entire 2D region of space set by the input box size.
We distinguish two key processes: first, a wave breaking that occurs independently along the two orthogonal dimensions and, second, the interaction of the generated solitons. 
In the early stages of the breaking, DSWs are generated at the box corners.
Then, the breaking proceeds along $x$ from the edges to the center of the box.
This dynamics strongly differs from the generation of a 1D SG by noise-seeded modulation instability, where solitons of variable amplitude and velocity emerge in random positions over the constant background~\cite{suret2023soliton}.
We identify the breaking geometry from the space region occupied by the DSWs generated at the boundary. The intensity profiles, averaged over designed regions of interest,
identify different rows along which the breaking occurs.
The number of breaking rows increases with the box height. 
From the profiles in Fig.\ref{fig:evolution}, two features emerge: 
the lack of breaking along $y$ in the 1D case 
and the random distribution of the peaks along $x$ in the 2D case. 
This non-periodic profile with localized modes is a key feature of a random set of solitons. 
By investigating the relation between the width and the peak intensity of each mode,
we prove the solitonic nature of the 2D localized modes through the observation of the soliton existence curve (see Supplementary Material).

\begin{figure}
    \centering
    \includegraphics[width=0.85\linewidth]{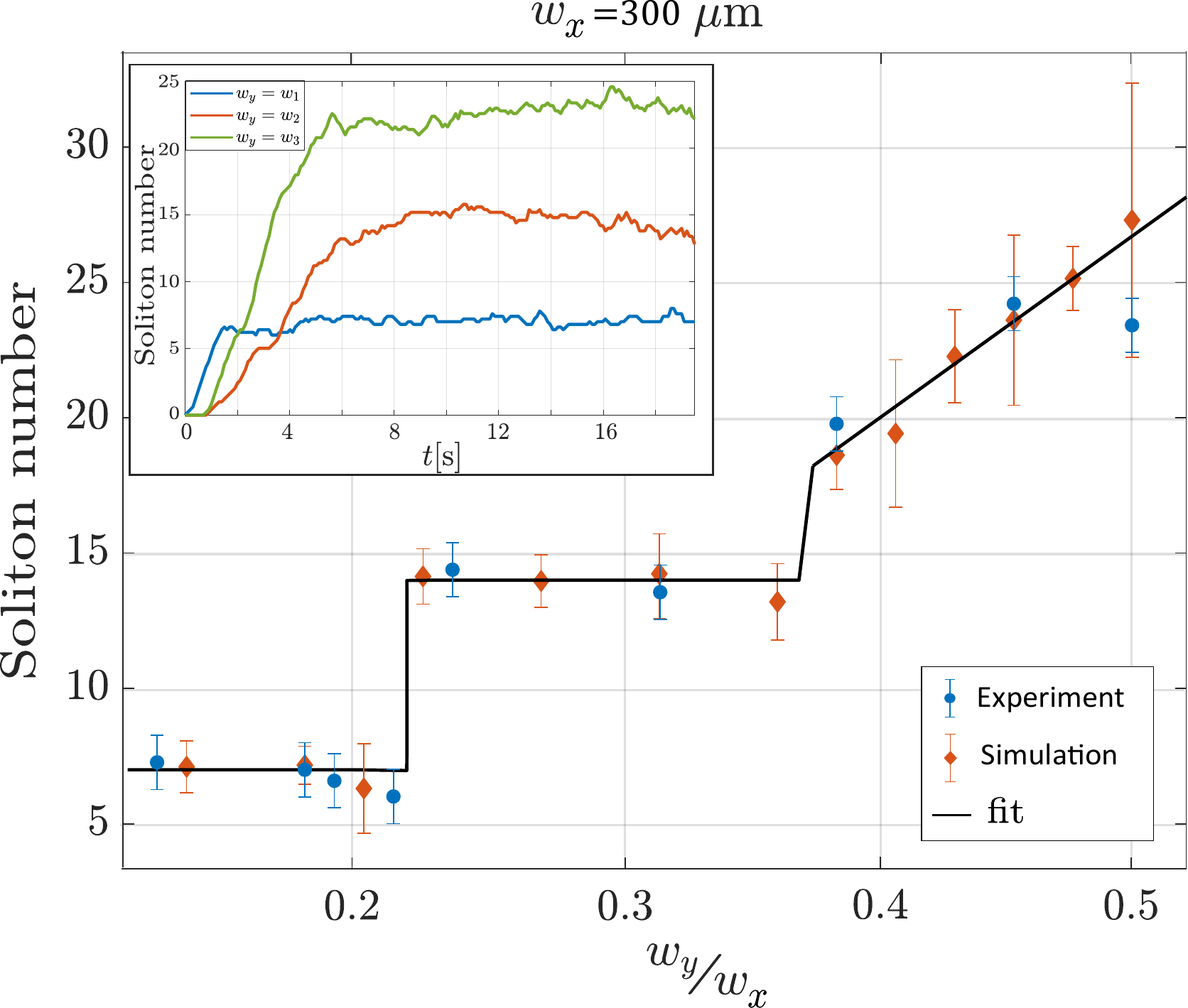}
    \caption{Number of solitons (mean) that emerge from the breaking of boxes of different heights
    in experiments and simulations. Numerical data are averaged over different noise realizations. Inset: soliton number as a function of the evolution time (experimental data). The plateau indicates a steady-state with a constant number of solitons.
    }
    \label{fig:soliton_number}
\end{figure}

To understand the collective solitonic phase, we evaluate the number of generated solitons. Due to the non-integrability of Eq.(\ref{eq:NLS_kerrsat}), the soliton number is expected to decrease during the evolution due to inelastic collisions. Differently, the number of solitons at different times after the formation of the ensemble presents only minor oscillations around a mean value. The measured evolution of the number of solitons is inset in Figure \ref{fig:soliton_number}.  Soliton fusions are not observed in the experimentally accessible times because the solitons propagate mainly in the forward direction (see Supplementary Fig. 2). The conservation of the number of solitons contrasts with a soliton-merging picture indicating the presence of a gaseous phase of 2D solitons. 

Figure \ref{fig:soliton_number} reports the soliton number for boxes of different heights and box width $w_x=300\mu$m. The soliton number depends on the total area of the box.
However, its variation by changing the aspect ratio furnishes fundamental information about the breaking process: (i) there is a sharp jump at the transition from 1D to 2D dynamics, with the number of solitons that doubles for $w_y/w_x\simeq1/4$; (ii) the breaking process occurs by rows, as demonstrated by an additional jump where the soliton number triples with respect to the 1D case;
(iii) the formation of breaking rows is observed as far as the side lengths of the box are significantly different. For $w_y/w_x\geq 0.4$, the soliton number increases linearly with the box area. 
This condition corresponds to a breaking process occurring simultaneously along $x$ and $y$.

\begin{figure*}
\centering
    \includegraphics[width=0.8\linewidth]{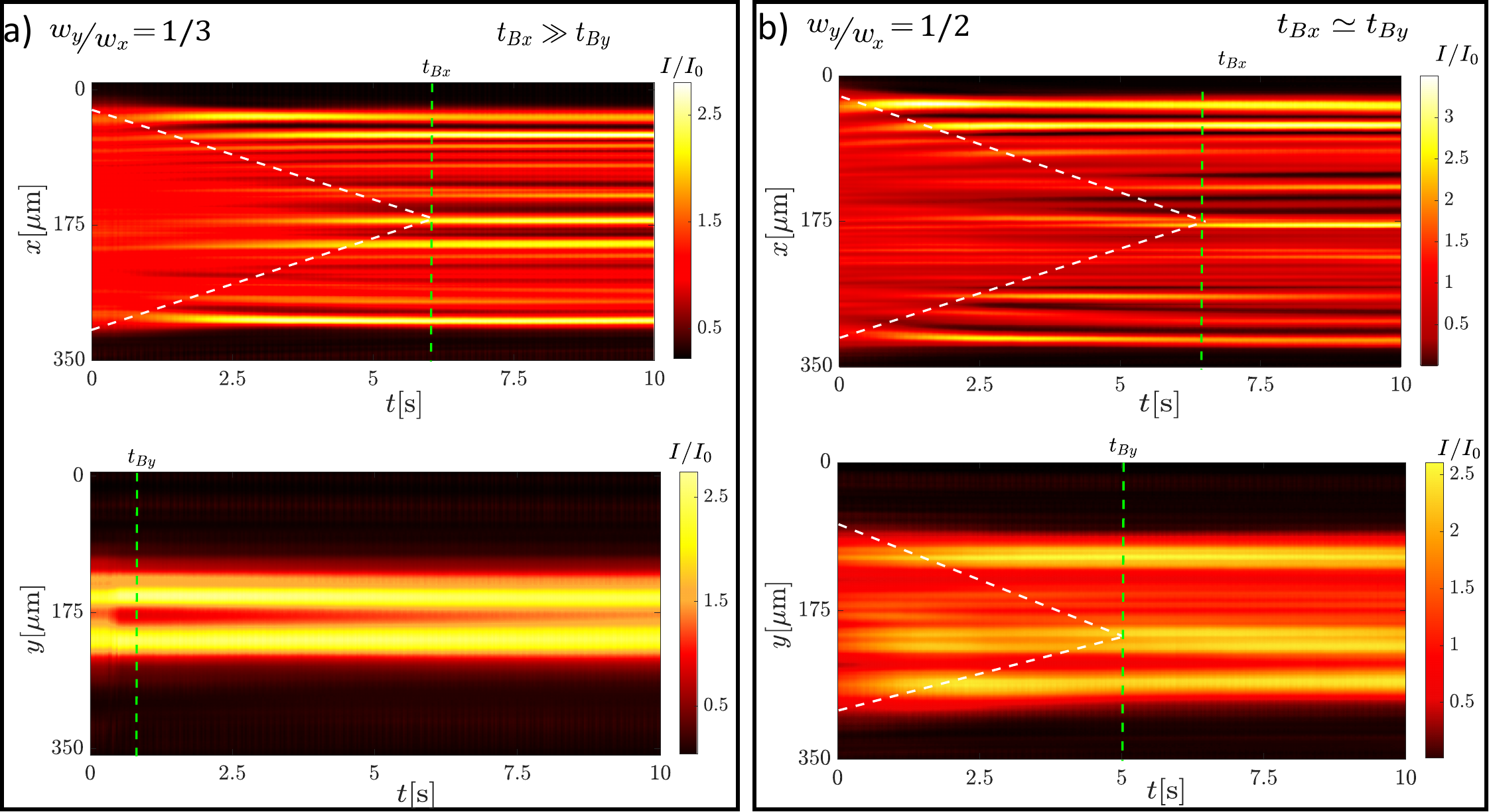}
    \caption{Experimental evolution of 2D boxes. (a) Intensity as a function of time for $w_y=w_2$.
     Along the $x$-axis is evident the formation and propagation of two DSWs. The green dashed line indicates the breaking time and the shock collision. White dashed lines show the calculated shock trajectories. The evolution along $y$ shows a breaking into two coherent localized structures at $t_{By}<t_{Bx}$. (b) Intensity evolution for $w_y=w_3$. Both sides of the box break up through shock collisions with comparable breaking times.}
    \label{fig:time_evolution}
\end{figure*}

The breaking of the box occurs independently in the two directions on a time scale proportional to the side length. 
When the lengths are different, the short side breaks much faster than the long side, causing the breaking by rows.
The experimental observation of this behavior is reported in Fig.\ref{fig:time_evolution}.
The aspect ratio $w_y/w_x$ determines the difference between the breaking timescales and the soliton number.
Fig.\ref{fig:time_evolution}(a) shows the time evolution of the box with $w_y=w_2$.
We define the breaking time $t_{Bx}$ as the instant of the shock collision along $x$.
We measure $t_{Bx}$ as the point where the emerging peak reaches the maximum intensity (collision point). 
We derive the propagation speed of the DSWs as $v_S= w_x/(2t_{Bx})$ ~\cite{marcucci2019topological}, 
which gives the shock trajectories shown in Fig.\ref{fig:time_evolution}(a).
The time evolution of the short side [Fig.\ref{fig:time_evolution}(a)] 
shows that the breaking does not involve DSWs but the formation of two coherent structures that propagate keeping their waveform. We define the corresponding breaking time $t_{By}$ as the instant of soliton formation.
 We find that $t_{Bx}>t_{By}$, resulting in the formation of two breaking rows.
Fig.\ref{fig:time_evolution}(b) shows the time evolution of the box with $w_y=w_3$.
Both sides break up through the collision of shocks with similar velocity. 
$t_{Bx}$ and $t_{By}$ have comparable values (simultaneous breaking).
The intensity evolution observed along each direction is consistent with the 
inverse-scattering solution of a 1D box \cite{jenkins2011semiclassical}, 
where the number of emerging solitons is proportional to the box width.

\begin{figure}
    \centering
    \includegraphics[width=0.9\linewidth]{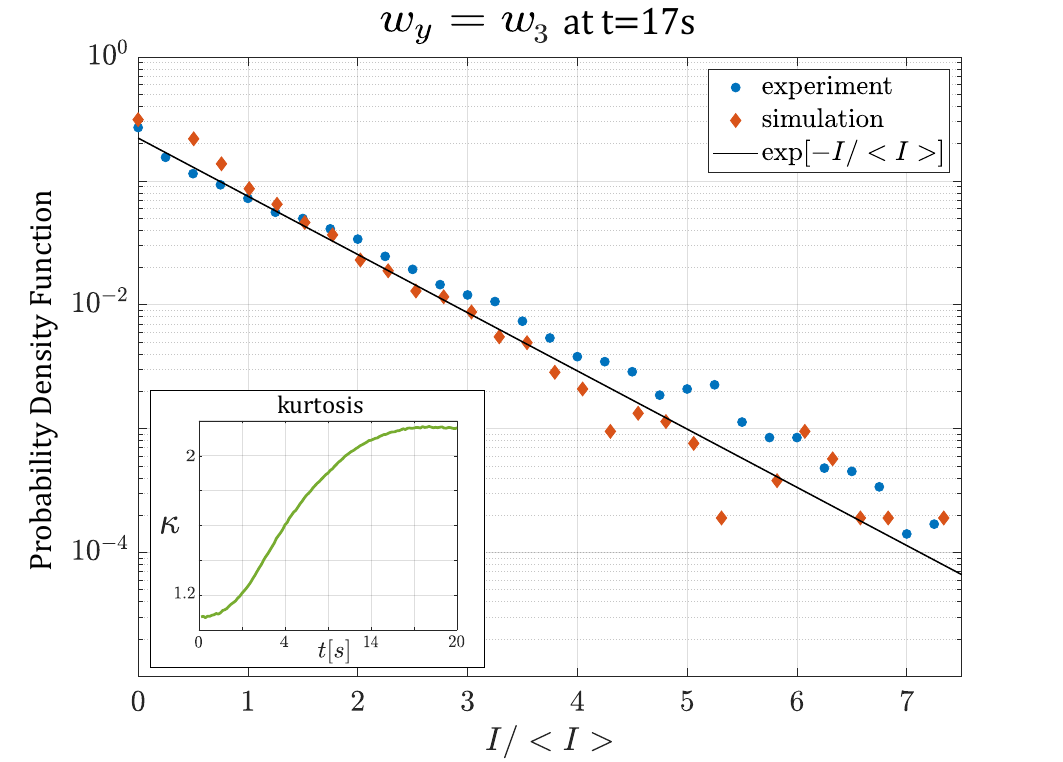}
    \caption{Intensity PDF of the gaseous phase of solitons. Experimental measurements for $w_y=w_3$ (blue dots), simulations (orange dots), and exponential fit (black line). The inset shows the measured time evolution of the kurtosis. }
    \label{fig:statistics}
\end{figure}

To characterize the gaseous phase of 2D solitons, we analyze its statistical properties. 
As evident in Fig.\ref{fig:evolution}, the field evolution is confined to the 2D region occupied by the input box, the gas volume. 
We evaluate the intensity statistics of the soliton ensemble in the gas volume for $w_y=w_3$.
The intensity PDF normalized to the mean intensity is reported in Fig. \ref{fig:statistics}.
Experimental and numerical data are well-fitted by a single exponential distribution.
This distinguishes our collective state from periodic structures originating from modulation instability.
An exponential PDF characterizes a variety of complex fields.
In the SG context, the exponential behavior is expected for a dense 1D SG with low velocity \cite{gelash2018strongly}. Our 2D soliton phase shows this feature. 
We report in Fig.\ref{fig:statistics} the measured kurtosis during the time evolution, which presents a significant increase and reaches a value close to the $\kappa= 2$  at the steady state as expected for an exponential PDF. 
The saturation of the kurtosis denotes the conservation of the soliton number, confirming the absence of soliton coalescence. 
We remark that the exponential PDF is spontaneously reached and not synthesized via the initial condition, suggesting a thermalization-like process during the beam evolution.
The statistical analysis supports the observation of a gaseous phase of 2D solitons with a PDF exhibited also by 1D SGs.

In conclusion, our investigation reveals the dynamics of a 2D dam break flow within a multi-dimensional photon fluid along with the spontaneous formation of a 2D gaseous phase of solitons.
The observation of the collision of orthogonal DSWs with distinct breaking points discloses the key role of dimensionality in the wave-breaking and forms the mechanism for the formation of the gaseous phase.
Through the controlled manipulation of the input wave, we have explored the transition from a 1D regime, characterized by solitons moving in a single direction, to a 2D scenario where solitons disperse and interact across the entire available space. The 2D nature can also introduce soliton fusions into intense waves \cite{xin2022intense}. Nevertheless, we have found the emergence of a gaseous phase characterized by a constant number of solitons.
Our findings suggest that the loss of integrability in two dimensions does not destroy the  
phenomenology of the SG, in analogy with the universality of other analytical soliton solutions \cite{tikan2017universality}. As a consequence, the work leaves an important open question about the possibility of defining a multi-dimensional SG in a non-integrable system.
From an application perspective, the complexity of the 2D SG can be exploited for enhanced functionalities in wave-based neuromorphic computing \cite{marcucci2020theory, silva2021reservoir, lopez2023self}.

We acknowledge support from the European Innovation Council (EIC) under the project HEISINGBERG (No. GA 101114978).


%

\end{document}


\title{Supplementary Material for: \\ \vspace*{0.1cm}
Observation of 2D dam break flow in a photon fluid}
\author{Ludovica Dieli}
\email{ludovica.dieli@uniroma1.it}
\affiliation{Department of Physics, Sapienza University, 00185 Rome, Italy}
\affiliation{Enrico Fermi Research Center, 00184 Rome, Italy.}
\author{Davide Pierangeli}
\affiliation{Institute for Complex Systems, National Research Council, 00185 Rome, Italy}
\author{Eugenio DelRe}
\affiliation{Department of Physics, Sapienza University, 00185 Rome, Italy}
\author{Claudio Conti}
\affiliation{Department of Physics, Sapienza University, 00185 Rome, Italy}


{ \let\clearpage\relax
\maketitle }
\subsection{Supergaussian beam shaping}
The box-shaped initial condition is prepared by shaping the optical field as a supergaussian with a flat phase. 
Letting the optical field $E(x,y,z)=A(x,y,z)e^{i\phi(x,y,z)}$, the input writes as 
\begin{equation}
  A(x,y,0)=A_0 \exp{\left[-\left(\frac{x}{w_{x}/2}\right)^{20}-\left(\frac{y}{w_{y}/2}\right)^{20}\right]} \, ,
    \label{eq:initial_condition}
\end{equation}
and $\phi(x,y,0)=0$. $w_{x,y}$ are the supergaussian widths, corresponding to the box width and height. 

\subsection{Saturable Kerr nonlinearity in photorefractive crystals}
In our experiment, nonlinear optical propagation occurs in a non-centrosymmetric photorefractive crystal, SBN61. The medium nonlinear response is due to a space charge field $E_{SC}$, generated by the combined effect of an external electrostatic field and the beam intensity. The space charge field generates an intensity-dependent variation in the refractive index $n(I,t)=n_0+\delta n(I,t)$ which is also time-dependent due to the charge dynamics inside the crystal. \\
According to the paraxial approximation of the optical field $E(x,y,z,t)=A(x,y,z)e^{(ikz-i\omega t)}$, the propagation is described by the well-known generalized 2D nonlinear Schroedinger equation
\begin{equation}
    2 {\textrm i} k\, \partial_{z}A+\left(\partial_{x}^{2}A+\partial_{y}^{2}A\right)+2 k^{2}\frac{\delta n(I,t)}{n_0} A=0
    \label{eq:NLS_kerrsat}
\end{equation}
where $k=k_0 n_0$, $k_0$ the vacuum wave-vector. \\
For a photorefractive crystal of tetragonal structure such as SBN, the electro-optic effect is linear and 
\begin{equation}
    \delta n=-\frac{1}{2}n_0^3 r_{33} E_{SC}.
\end{equation}
where $\boldsymbol{r}$ is the crystal electro-optics tensor, and $r_{33}$ is the only relevant component because of the direction of the applied field.\\
The space charge field is time-dependent, \cite{delre2009photorefractive},  and described by 
\begin{equation}
    \frac{\partial E_{SC}}{\partial t}=\frac{E_{\text{ext}}}{t_d}-\frac{E_{SC}}{t_d}\left(1+\frac{I}{I_b}\right)
\end{equation}
where $E_{\text{ext}}=V/L_x$, $V$ applied voltage, and $L_x$ distance between the $2$ electrodes, $I_b$ is the background intensity and $t_d$ is the characteristic dielectric time constant, depending on the material recombination rate, the impurity densities, and the background intensity. \\
Neglecting for simplicity time nonlocality (i.e. $\partial_t (1+I/I_b)\simeq0$), the equation has the solution 
\begin{equation}
    E_{SC}(I,t)=\frac{E_{\text{ext}}}{1+\frac{I}{I_b}}\left[ 1-e^{-\left(1+\frac{I}{I_b}\right) t/t_d}\right]=\frac{E_{\text{ext}}}{1+\frac{I}{I_b}}\left[ 1-e^{- t/\tau_s}\right]
    \label{ESC_time}
\end{equation}
where $\tau_s=t_d/(1+I/I_b)$ is the saturation time.\\
Thus, we get the saturable Kerr nonlinear equation:
\begin{equation}
     2ik\frac{\partial A}{\partial z}+\left(\partial_{x}^{2}A+\partial_{y}^{2}A\right)
+2k^2\frac{\delta n_0}{n_0}\frac{f(t)}{\left(1+\frac{I}{I_b}\right)}   A=0.
\label{eq:model_phys}
\end{equation}
 with $\delta n_0=-n_0^3r_{33}V/(2L_x)$ and response function $f(t)=(1-e^{-t/\tau_s})$.

\subsection{Generality of the observed 2D dam break flow}
We observed the 2D breaking dynamics of a box-shaped wave and the generation of a gaseous phase of solitons. Numerical simulations confirm the observations. Results refer to a saturable nonlinearity in the presence of noise. In order to support the generality of our results and confirm that the phenomenology also occurs for other kinds of nonlinearity, we investigate through numeric simulations. The results, reported in Supplementary Fig.\ref{fig:varie_NL}, confirm a breaking picture characterized by colliding DSWs along orthogonal directions even for a different nonlinearity. The investigated cases are the centrosymmetric photorefractive crystals, Ref.\cite{marcucci2019topological}, whose electro-optic response is quadratic, and a saturable nonlocal nonlinearity studied in Ref.\cite{delre2005mechanisms}.
\begin{figure}
     \centering
     \includegraphics[width=\linewidth]{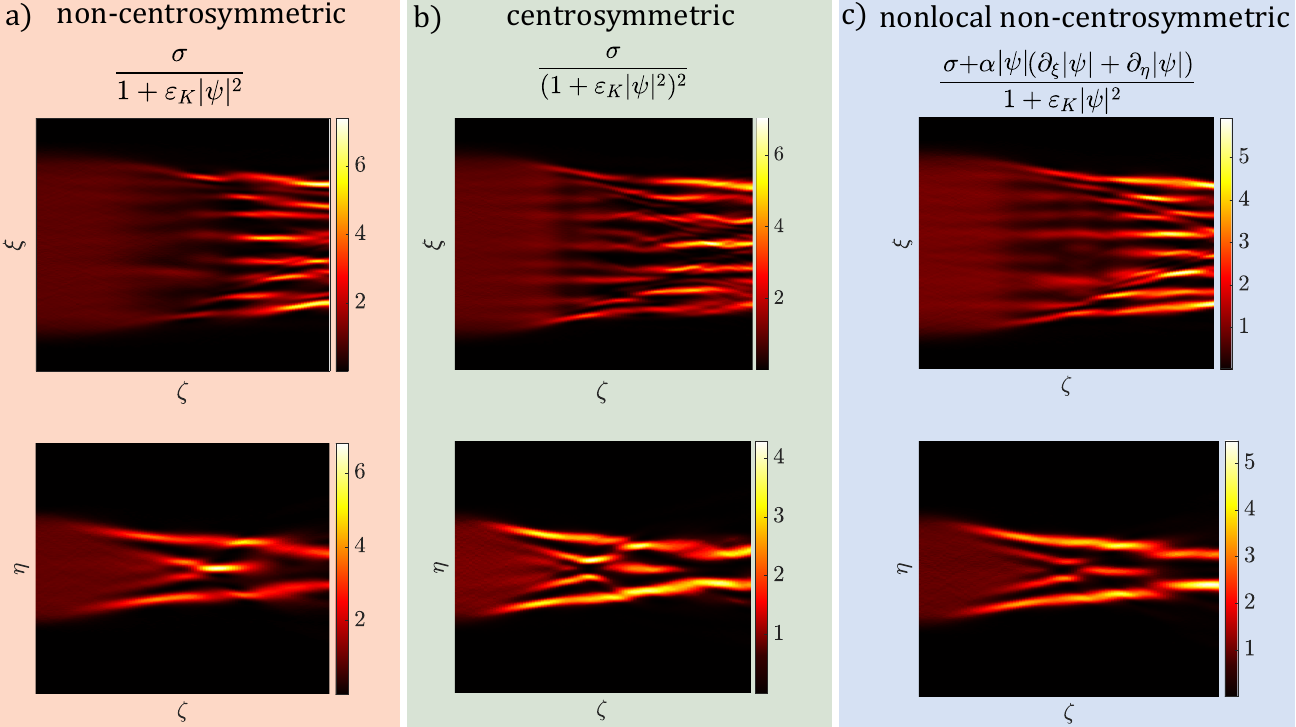}
     \caption{Nonlinear propagation of 2 box with $w_y=w_2$ simulated with different types of nonlinearity.  a) Saturable Kerr nonlinearity describing non-centrosymmetric photorefractive crystals. b) Saturable Kerr nonlinearity in centrosymmetric photorefractive crystals. c) Saturable nonlocal nonlinearity obtained in presence of higher-order effects. Expressions for the nonlinearity refer to the nonlinear term in Eq.\ref{eq:model_adim}.}
     \label{fig:varie_NL}
\end{figure}

\subsection{Mapping between time evolution and space propagation}
To model the experimental measurements, we introduce 
a dimensionless version of the equation that includes the time dependence in the propagation variable. 
By defining 
\begin{equation}
      \psi=\frac{E}{E_0},\quad \zeta=\frac{2z}{\varepsilon(t)z_D}, \qquad \xi=\frac{2x}{w_0}, \qquad \eta=\frac{2y}{w_0}; 
\end{equation}
where $w_0$ is an arbitrary length related to the beam width, and $z_D=\pi n_0 w_0^2/\lambda$ is the diffraction length of a Gaussian beam of width $w_0$. We obtain the equation
\begin{equation}
    i \varepsilon (t) \frac{\partial \psi}{\partial \zeta}+\frac{\varepsilon^2(t)}{2} \left[\frac{\partial^2 \psi}{\partial \xi^2}+\frac{\partial^2 \psi}{\partial \eta^2}\right]+\frac{1}{1+\varepsilon_K|\psi|^2} \psi=0.
    \label{eq:model_adim}
\end{equation}
The time dependence is now encoded in the propagation variable $\zeta$ through the parameter $\varepsilon(t)$
\begin{equation}
    \varepsilon(t)=\frac{w_0}{z_D}\sqrt{\frac{n_0}{\delta n_0 f(t)}}.
\end{equation}
Thus, the box evolves in time and propagates in $z$ simultaneously and the field variation can be written as
\begin{equation}
    \frac{\partial \psi}{\partial \zeta}=\frac{\partial z}{\partial \zeta}\frac{\partial \psi}{\partial z}+\frac{\partial t}{\partial \zeta}\frac{\partial \psi}{\partial t}.
    \label{eq:derivative_propagation_variable}
\end{equation}

\begin{figure}
    \centering
    \includegraphics[width=0.8\linewidth]{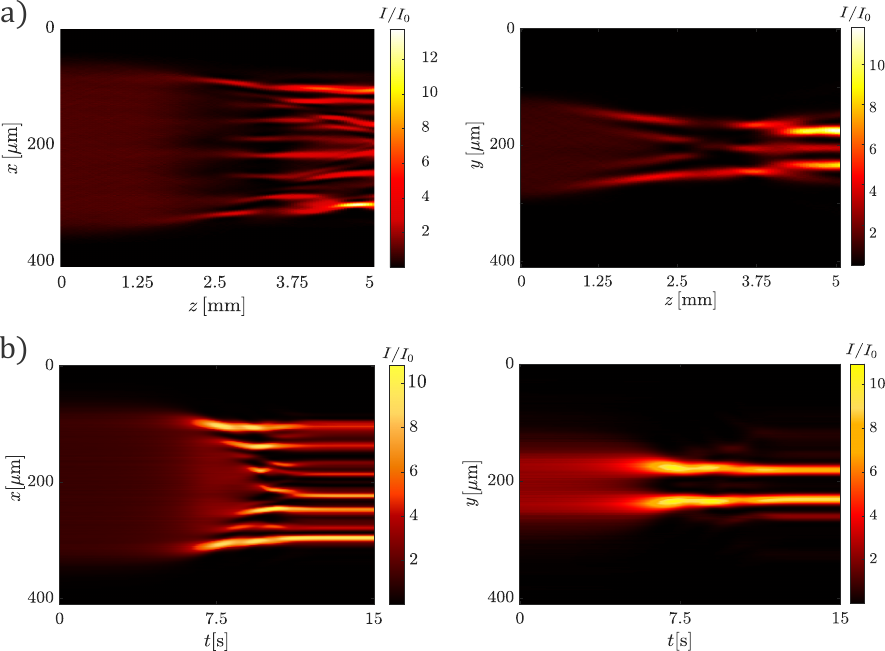}
    \caption{Space propagation and time evolution of the optical field. a) Simulation of space propagation at a fixed time. b) Simulation of time evolution at the final propagation distance $z=z_{\text{fin}}=5$mm.}
    \label{fig:compare_zeta_time}
\end{figure}

However, in the experiment, we observe the output facet of the crystal, corresponding to the fixed value of $z=z_{\text{fin}}$, and the variation of the field $\psi$ in the propagation variable $\zeta$, eq.(\ref{eq:derivative_propagation_variable}) reduces to a time variation.\\
The simulations have been realized as $z$-propagation at a fixed time, considering $\varepsilon(t)$ a constant. The two methods give comparable results, illustrated by the example in Fig.\ref{fig:compare_zeta_time}.
Note that the differences between space propagation and time evolution are due to the initial condition. Indeed, in the time evolution, at $t=0$ and  $z=z_{\text{fin}}$ we observe the linear propagation of the box, while in the $z$-propagation, at $z=0$ the box is unevolved. Diffraction affecting the  linear propagation, most evident for the short side, causes the small differences between the two cases. However, the steady states, reached after the breaking, are comparable.

\subsection{Soliton existence curve}
\begin{figure}
    \centering
    \includegraphics[width=0.5\linewidth]{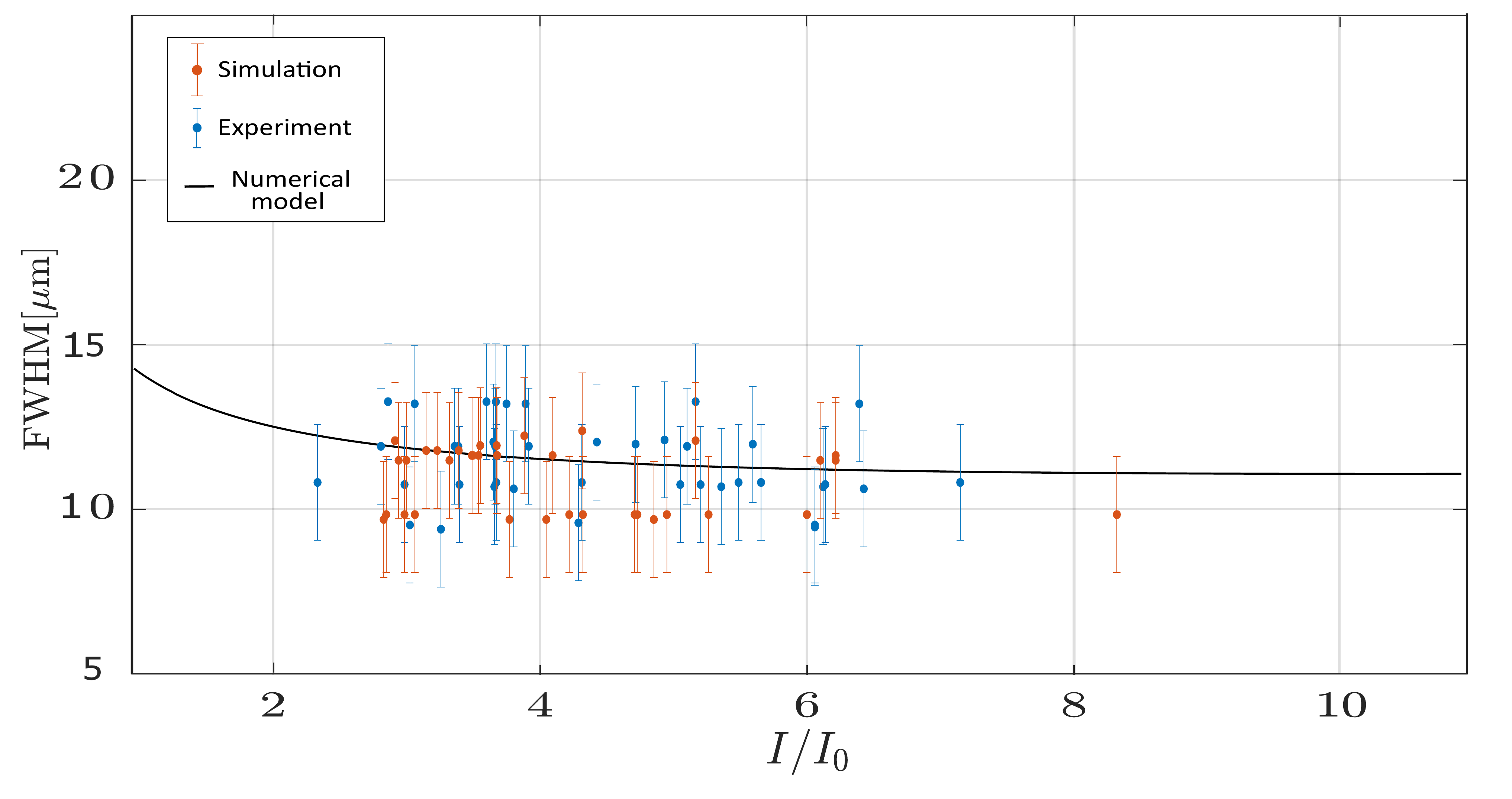}
    \caption{Soliton existence curve. Relation between the FWHM of the solitons and their peak intensity. Orange dots are simulated data, blue dots are experimental data, and the black line is the curve obtained by numerical analysis.} 
    \label{fig:existence_curve}
\end{figure}

The nonlinear evolution of the box-shaped optical field results in the generation of an ensemble of 2D solitary waves. The proof of the solitonic nature of the observed localized modes can be found in the soliton existence curve which relates the soliton width to its peak intensity. Defining the mode width as the full-width half maximum (FWHM) of the intensity distribution around each peak, we measure the FWHM of the generated 2D solitons in experiments and simulations, obtaining the existence curve reported in Fig.\ref{fig:existence_curve}.\\
We compare the data with the existence curve obtained by numerical integration of the model equation.
The model equation is not integrable, but, solitonic solutions can be numerically found by assuming radial symmetry\cite{crosignani1997three}. The resulting curve provides the relation between the dimensionless width of the solitonic solutions and the normalized intensity of the peak. The dimensionless width is related to the measured FWHM by a scaling parameter.
Supplementary Fig.\ref{fig:existence_curve} reports the result of the analysis. The error on the experimental points takes into account the $(x,y)$ asymmetry of the data due to the crystal anisotropy.
We note that 2D solitons in both experiments and simulations are well described by the expected curve. 


%